\documentclass[pra,aps,twocolumn,showpacs,superscriptaddress,floatfix,amsmath,amssymb,nofootinbib]{revtex4}
\usepackage{amsfonts}
\usepackage{graphicx}
\newcommand{\beq}{\begin{equation}}
\newcommand{\eeq}{\end{equation}}

\newcommand{\beqa}{\begin{eqnarray}}
\newcommand{\eeqa}{\end{eqnarray}}
\newcommand{\beqax}{\begin{eqnarray*}}
\newcommand{\eeqax}{\end{eqnarray*}}

\newcommand{\bZ}{{\mathbb{Z}}}
\newcommand{\bR}{{\mathbb{R}}}
\newcommand {\fabs}[1] {\left| #1 \right|}
\newcommand {\fexp} [1] {\exp \left( #1 \right)}

\newcommand {\fabsq}[1] {\left| #1 \right|^2}
\def\ra{\rangle}
\def\la{\langle}
\begin{document}
\title{Fast and stable manipulation of a charged particle in a Penning trap}
\author{A. Kiely}
\email{anthony.kiely@umail.ucc.ie}
\affiliation{Department of Physics, University College Cork, Cork, Ireland}
\author{J. P. L. McGuinness}
\affiliation{School of Mathematical Sciences, University College Cork, Cork, Ireland}
\author{J. G. Muga}
\affiliation{Departamento de Qu\'{\i}mica F\'{\i}sica, UPV/EHU, Apdo
644, 48080 Bilbao, Spain}
\affiliation{Department of Physics, Shanghai University, 200444
Shanghai, People's Republic of China}
\author{A. Ruschhaupt}
\affiliation{Department of Physics, University College Cork, Cork, Ireland}
\begin{abstract}
We propose shortcuts to adiabaticity which achieve fast and stable
control of the state of a charged particle in an electromagnetic field. In particular
we design a non-adiabatic change of the magnetic field strength in a Penning
trap which changes the radial spread without final excitations.
We apply a streamlined version of the fast-forward formalism as well as
an invariant based inverse engineering approach. We compare both methods and
examine their stability.
\end{abstract}
\maketitle
%
%
%
\section{Introduction}
The coherent manipulation of quantum systems for quantum technologies, fundamental studies,  or 
metrology  often requires control protocols of external parameters  
that are fast and stable with respect to  perturbations.  
This has motivated the development of shortcuts to adiabaticity (STA), which are 
schemes that reach the fidelities of adiabatic processes in shorter times, keeping or even 
enhancing their stability;
shortcuts to adiabaticity have been reviewed in \cite{sta_review}.

The fast expansion or compression of a particle state driven by a time dependent trap frequency 
is one of the paradigmatic operations for which STA have been developed, both in theory \cite{Ch10} and experiment \cite{Nice10,Nice11}.
Interacting particles in an expanding/compressing external harmonic potential  such as ion chains \cite{inprep}, 
Bose-Einstein condensates \cite{BEC}, or classical gases \cite{Boltzmann}, have also been studied.   
There are many applications for fast expansion/compression, such as controlled cooling/heating of the state \cite{Lianao,Lianao2}, implementation of 
quantum engines and refrigerators based on cyclic expansions and compressions \cite{Salamon09,EPL11}, 
fast switching between manipulations suited for a low trap frequency or for a high 
trap frequency configuration \cite{inprep}, or efficient sympathetic cooling \cite{Onofrio11,Onofrio12}. Fundamental aspects such as the quantification of the 
third law of thermodynamics \cite{Salamon09,energy} have also been examined.     

Real traps are of course three-dimensional, but most of the theory work deals with 1D traps with time dependent 
frequencies whose effective realization is not straightforward. Torrontegui et al. \cite{3d} studied  the      
fast expansions of cold atoms in a three-dimensional Gaussian-beam optical trap. The radial and axial frequencies are coupled and 
as a consequence some shortcut schemes that work in 1D were in fact restricted to certain parameter domains, 
and others failed completely. Traps with uncoupled radial and axial frequencies are of interest to perform clean  STA expansions/compressions and such a possibility is indeed provided by the Penning trap.  

In this paper, we will put forward shortcut schemes to control a charged particle in a Penning trap. 
A Penning trap uses a combination of a uniform
and unidirectional magnetic field and an electrostatic quadrupole potential. This potential
is typically created using three electrodes which are hyperboloids of revolution. Penning traps
are commonly used for accurate measurement of the properties of different charged particles \cite{werth2010}.

Since different operations on the trapped charged particle (preparation,
measurement, or interactions) may require or benefit from different extensions
of the density cloud, our aim is to change this extension rapidly without
producing final excitations.
Therefore, we shall construct schemes to
decrease the radial extension of the particle's wave function, without producing final excitations, by changing 
the magnetic field intensity. 

We shall first design such shortcuts by means of the fast-forward formalism.
The basic fast-forward formalism for a particle in a time dependent potential 
(without an electromagnetic field) was first
introduced by S. Masuda and K. Nakamura
\cite{masuda2008,masuda2010}. Later a streamlined version of this formalism was
developed in \cite{torrontegui2012}. This streamlined formalism has been applied for example
to engineering of fast and stable splitting of matter waves
\cite{torrontegui2013} and to achieve rapid loading of a Bose-Einstein condensate into an
optical lattice \cite{campo2014}. Another strategy to design shortcut schemes
is based on Lewis-Riesenfeld invariants \cite{LR69, Ch10}. Applications of
this strategy are reviewed in \cite{sta_review}.

In the following section we will develop a streamlined version of the
fast-forward formalism in the general case of a charged particle in an
electromagnetic field.
In Section \ref{sect_unidir}, we will review the known eigenstates of a particle in a 
a Penning trap. In section \ref{vary_field}, we apply the streamlined formalism
to change the state from
an eigenstate of one magnetic field strength to that of a larger field strength.
We also compare it with an inverse engineering approach based on Lewis-Riesenfeld invariants
and examine the stability versus systematic errors in the magnetic field.
Finally, in Section \ref{Discussion} we discuss our results.
%
%
%
%
\section {General streamlined formalism \label{sect_formalism}}
A fast-forward formalism including an electromagnetic field was introduced in \cite{masuda2011}.
We shall now put forward a streamlined version.
\subsection{Main equations}
We consider a spinless charged particle as spin will not play any role in the trap configuration
considered below. The Schr\"odinger equation for this particle in an electromagnetic field
is given by 
\beqa
i\hbar\frac{\partial}{\partial t} \Psi &=& H(t) \Psi,
\label{schr}
\eeqa
where the Hamiltonian (expressed in some chosen gauge e.g. the Coulomb gauge) is given
in coordinate representation by
\beqa
H(t) = \frac{1}{2m}
\left(\frac{\hbar}{i}\nabla - q \vec A(t, \vec r)\right)^2 + q\phi(t, \vec r),  
\eeqa
with $q$ being the charge, $\vec A$ the vector potential and $\phi$ the
scalar potential (both real).
We write $\Psi$ as
\beq
\Psi (t, \vec r) = \alpha(t, \vec r) e^{i \beta(t, \vec r)},
\label{ansatzpsi}
\eeq
where $\alpha(t, \vec r), \beta(t, \vec r) \in \mathbb{R}$. Note that $\Psi
(t, \vec r)$ corresponds to the fast-forwarded state
$\Psi_{FF}$ in \cite{masuda2011}. $\vec A$ and $q\phi$ correspond to the driving potentials
$\vec A_{FF}$ and $V_{FF}$ in \cite{masuda2011}. 

Inserting the ansatz \eqref{ansatzpsi} into Eq. \eqref{schr} and then multiplying the equation by
$e^{-i \beta(t, \vec r)}$,  we get for the real part of the result
\beqa
0 = -\frac{\hbar^2}{2m} \Delta \alpha + \frac{1}{2m} \left(q\vec A - \hbar
\nabla \beta\right)^2 \alpha + \left(q \phi + \hbar \frac{\partial
  \beta}{\partial t}\right) \alpha,
  \nonumber\\
\label{re}
\eeqa
and for the imaginary part
\beqa
\hbar \frac{\partial\alpha}{\partial t}
= \frac{\hbar}{2m}\nabla\left(q\vec A - \hbar\nabla \beta\right) \alpha
+ \frac{\hbar}{m}\left(q\vec A - \hbar\nabla \beta\right) \nabla \alpha.
\label{im}
\eeqa
To write these two equations in a more compact way, let us define
\beqa
\vec\chi := \vec A - \frac{\hbar}{q} \nabla \beta, \;
\Phi := \phi + \frac{\hbar}{q} \frac{\partial \beta}{\partial t}.
\label{newvar}
\eeqa
The electric and the magnetic fields are now given by
\beqa
\vec E = -\nabla \Phi - \frac{\partial \vec\chi}{\partial t},\;
\vec B = \nabla \times \vec\chi.
\label{eb}
\eeqa
Using these definitions of $\vec\chi$ and $\Phi$,
the two equations (\ref{re}) and (\ref{im}) simplify to
\beq
\Phi = \frac{\hbar^2}{2 m q \alpha} \Delta \alpha - \frac{q}{2m} \vec\chi^2,
\label{main_eq_phi}
\eeq
and
\beqa
\frac{\partial\alpha}{\partial t}
- \frac{q}{2m}(\nabla\vec\chi) \alpha
- \frac{q}{m} \vec\chi \nabla \alpha = 0.
\label{main_eq_chi}
\eeqa
These are the two main equations.

Note that $\vec\chi$ and $\Phi$ as well as the main equations
(\ref{main_eq_phi}) and (\ref{main_eq_chi}) are invariant under a gauge transformation $\Lambda$ acting in
the usual way
\beqa
\vec A \to \vec A + \nabla\Lambda, \;
\phi \to \phi - \frac{\partial\Lambda}{\partial t}\; \mbox{and}\;
\Psi &\to& e^{\frac{i}{\hbar}q\Lambda}\Psi,
\eeqa
i.e., $\beta \to \beta + \frac{q}{\hbar}\Lambda$ and $\alpha$ is unchanged.
\subsection{Inverse engineering and boundary conditions}
Let the initial state of the system
$\psi_0 (\vec r) \equiv \alpha_0(\vec r) e^{i \beta_0(\vec r)}$
be an eigenstate of the initial time independent Hamiltonian
\beqa
H_0 &=& \frac{1}{2m} \left(\frac{\hbar}{i}\nabla - q \vec A_0\right)^2 + q\phi_0,
\eeqa
with eigenvalue $ \mathcal{E}_0$. (The eigenstates of the Penning trap are reviewed in the following section.)
The goal is to design a scheme ($\vec A(t, \vec r)$ and $\phi(t, \vec r)$)
such that
the final state (at $t=T$) of the system, $\psi_T (\vec r) \equiv
\alpha_T(\vec r) e^{i \beta_T(\vec r)}$, is an eigenstate of the final (time independent) Hamiltonian
\beqa
H_T &=& \frac{1}{2m} \left(\frac{\hbar}{i}\nabla - q \vec A_T\right)^2 +
q\phi_T,
\eeqa
with eigenvalue $\mathcal{E}_T$. 
The Hamiltonian should be continuous at initial and final time, i.e.,
$H (0) = H_0$ and $H(T) = H_T$.

In the inversion protocol we first design $\alpha(t, \vec r)$ and $\beta(t, \vec r)$
fulfilling the boundary conditions
\beqa
\alpha (0,\vec r) = \alpha_0 (\vec r), \; \alpha (T, \vec r) = \alpha_T (\vec r),
\nonumber\\
\beta  (0,\vec r) = \beta_0 (\vec r), \; \beta (T, \vec r) = \beta_T (\vec r).
\label{bound_ab}
\eeqa
In the next step, we have to solve for $\vec\chi$ in Eq. (\ref{main_eq_chi}). The
function $\Phi$ is then given by Eq. (\ref{main_eq_phi}).
Because the Hamiltonian should be changing continuously at initial and final
time, $\vec\chi$ must fulfill the following boundary conditions
\beqa
\vec\chi(0, \vec r) = \vec A_0 - \frac{\hbar}{q} \nabla\beta_0, \;
\vec\chi(T, \vec r) = \vec A_T - \frac{\hbar}{q} \nabla\beta_T.
\label{bound_chi}
\eeqa
A consequence of these conditions can be seen by evaluating
Eq.(\ref{main_eq_chi}) at the initial and final time (see also
Appendix \ref{app_boundary}). This leads to
\beq
\frac{\partial\alpha}{\partial t} (0,\vec r) = 0, \;
\frac{\partial\alpha}{\partial t} (T,\vec r) = 0.
\eeq
The boundary conditions of $\Phi$ can be seen by evaluating
Eq. (\ref{main_eq_phi}) at initial and final time leading to (see also
Appendix \ref{app_boundary})
\beqa
\Phi(0,\vec r) = \phi_0 (\vec r) - \frac{1}{q} \mathcal{E}_0 = \phi_0 (\vec r)
+ \frac{\hbar}{q} \frac{\partial\beta}{\partial t} (0,\vec r),
\nonumber \\
\Phi(T,\vec r) = \phi_0 (\vec r) - \frac{1}{q} \mathcal{E}_T = \phi_0 (\vec r)
+ \frac{\hbar}{q} \frac{\partial\beta}{\partial t} (T,\vec r).
\eeqa 
These conditions are equivalent to
\beqa
\frac{\partial\beta}{\partial t} (0,\vec r) = -\frac{1}{\hbar} \mathcal{E}_0,\;
\frac{\partial\beta}{\partial t} (T,\vec r) = -\frac{1}{\hbar} \mathcal{E}_T.
\label{bound_beta}
\eeqa
Finally, the vector potential and the scalar potential in the
chosen gauge are then given by
\beqa
\vec A (t, \vec r) &=& \vec\chi(t,\vec r) + \frac{\hbar}{q} \nabla\beta(t,\vec
r),
\\
\phi(t,\vec r) &=& \Phi(t,\vec r) - \frac{\hbar}{q}
\frac{\partial\beta}{\partial t},
\eeqa
and the electric and magnetic fields are given by Eq. (\ref{eb}).

The above boundary conditions guarantee that the magnetic field is
continuous at the initial and final time. 
To make the electric field continuous at the initial and final time, we also
impose
\beqa
\frac{\partial\vec\chi}{\partial t} (0,\vec r) = 0, \;
\frac{\partial\vec\chi}{\partial t} (T,\vec r) = 0.
\label{bound_electric}
\eeqa
%
%
%
%
%
%
\section{Energy Eigenstates of a Penning trap \label{sect_unidir}}
Let us introduce cylindrical coordinates $\{r,\theta,z\}$ where
$x=r\cos\theta$ and $y=r\sin\theta$, and define
the orthogonal unit vectors
\beqa
\hat r = \left(\begin{array}{c} \cos\theta\\ \sin\theta\\0 \end{array}\right), \; 
\hat \theta = \left(\begin{array}{c}
  -\sin\theta\\ \cos\theta\\0 \end{array}\right), \;
  \hat z = \left(\begin{array}{c}
  0\\0\\1 \end{array}\right).
\eeqa
For the Penning trap, we assume a homogeneous magnetic field in $z$ direction, $\vec B = B_z \hat z$,
and an electrostatic field of the form $\vec E = E_r \hat r + E_\theta \hat\theta + E_z \hat z$,
where
\begin{eqnarray}
E_r = \frac{m \omega_z^2}{2q} r, \; E_\theta = 0, \; E_z = - \frac{m \omega_z^2}{q} z.
\end{eqnarray}
The vector potential and the scalar potential can be written as 
\begin{equation}
\vec{A}= \frac{r B_z}{2} \hat\theta, \;
 \phi=\frac{m \omega_{z}^2}{4 q}\left(2z^2-r^2\right).
\end{equation}
The corresponding Hamiltonian  reads
\begin{eqnarray}
H = - \frac{\hbar^2}{2m} \Delta +
\frac{m}{2} \widetilde{\omega}^2 r^2 - \omega L_z +\frac{1}{2} m \omega_z^2 z^2,
\label{statham}
\label{LRHam}
\end{eqnarray}
where $\widetilde{\omega}^2=\omega^2-\omega_{z}^2/2$, $\omega = q B_z/(2m)$ and 
$L_z = \frac{\hbar}{i} \frac{\partial}{\partial \theta}$ is the $z$-component of the
angular momentum operator. $L_z$ commutes with the rest of the Hamiltonian, so it represents a conserved
quantity. 
The Hamiltonian \eqref{LRHam} is separable into a Hamiltonian depending on $r$
and $\theta$ and a Hamiltonian depending solely on $z$.
The corresponding Schr\"odinger equation can be solved by a product of a
function of $r$ and $\theta$
and a function of $z$.
The $z$-dependent function describes the dynamics of a harmonic oscillator with axial (angular) frequency $\omega_z$ which 
we assume to have a fixed value. Since we shall  
consider $B_z$ as the time dependent external parameter, the non-trivial part of 
interest is the function that depends on $r$ and $\theta$. Hence we focus solely on this part. We also assume $q B_z > 0$.
The energy eigenfunctions are
\beq
\psi_{N,M,l}(r,\theta) = f_{N,M,l}(r) \fexp{iM\theta},
\eeq
where
\begin{eqnarray}
f_{N,M,l}(r)&=&\frac{1}{\sqrt{2\pi}}\sqrt{\frac{N!}{(N+|M|)!}}\frac{1}{l}\left[\frac{r}{\sqrt{2}l}\right]^{|M|}
\nonumber\\
& & \times \fexp{-\frac{r^2}{4l^2}}
 L^{|M|}_N\left(\frac{r^2}{2l^2}\right),
\end{eqnarray}
$N, M \in \bZ$ , \(L^{|M|}_N (q)\) are the generalized Laguerre Polynomials, defined by
\begin{equation}
L^{a}_N (Q)=\frac{Q^{-a}e^Q}{N!} \frac{d^N}{dQ^N}(e^{-Q}Q^{N+a}),
\end{equation}
and the constant \(l\) is defined by
\begin{equation}
l=\sqrt{\frac{\hbar}{2 m \widetilde{\omega}}}.
\label{eq_l}
\end{equation}
$M$ is the quantum number associated with the $z$ component of the angular
momentum operator (i.e. $L_{z} \psi_{N,M,l}(r,\theta)= M \hbar
\psi_{N,M,l}(r,\theta)$ ) and $N$ is a quantum number that determines the radial
structure. $l \in \bR$ is the characteristic radial length scale of the
wavefunction; it is determined by the magnetic field $B_z$ and the axial
frequency $\omega_z$ via Eq. \eqref{eq_l}.
The energy eigenvalues are
\beq
\mathcal{E}_{N,M} = \hbar\widetilde{\omega} \left(2N + |M|+ 1\right)-\hbar \omega M,
\eeq
where $N=0,1,...$ and $M$ is an integer.
Alternatively, using $\widetilde N := 2N + \fabs{M}$, the energy eigenvalues are often written as
$\mathcal{E} = \hbar \widetilde{\omega} (\widetilde N+1) - \hbar\omega M$ with
$\widetilde N=0,1,....$ and $M=-\widetilde N, -\widetilde N+2,...,\widetilde N-2,\widetilde N$.
For example, these eigenfunctions were previously found
for $\widetilde{\omega}=\omega$ in \cite{brito2007}.
%
%
%
%
\section{Varying the magnetic field strength \label{vary_field}}
We would like to design the time dependence of the magnetic field so that 
the system starts from
the eigenstate $\psi_{N,M,l_0}(r,\theta)$ at initial time $t=0$ with magnetic
field $B_z(0)=B_{z,0}$ and ends in
the eigenstate  $\Psi_{N,M,l_T}(r,\theta)$ at final time $t=T$ with magnetic
field $B_z(T)=B_{z,T}$, i.e.
$\Psi (0,r,\theta) = \psi_{N,M,l_0}(r,\theta)$,
$\Psi (T,r,\theta) = \psi_{N,M,l_T} (r,\theta)$,
where $l_0 = \sqrt{\hbar/(2 m \widetilde{\omega}(0))}$,
$l_T = \sqrt{\hbar/(2 m \widetilde{\omega}(T))}$,
and $\widetilde{\omega}(t)=\sqrt{\left(q B_z(t)/(2m)\right)^2 - \omega_{z}^2/2}$.
The frequency $\omega_z$ should be kept constant.
Of course, this can be done in an adiabatic way but here we want to derive a
shortcut to adiabaticity. As an example, we will examine a way to decrease 
the characteristic length scale,
$l_0 \to l_T$. 
\subsection{Streamlined formalism}
Following the algorithm presented in Section \ref{sect_formalism}, we start by
choosing the following ansatz for the time evolution of the wavefunction
\begin{eqnarray}
\alpha (t,r) &=& \sqrt{\frac{N!}{(N+\fabs{M})!}} \frac{1}{\sqrt{2\pi} l(t)}
\left(\frac{r}{\sqrt{2} l(t)}\right)^{\fabs{M}}\nonumber\\
&& \times \fexp{-\frac{r^2}{4 l(t)^2}} L_N^{\fabs{M}} \left(\frac{r^2}{2 l(t)^2}\right),
\label{alpha}
\end{eqnarray}
and $\beta (t,\theta) = M \theta + \zeta (t)$.
For the boundary conditions \eqref{bound_ab}, it follows $\alpha (0,r) = f_{N,M,l_0} (r)$ and
$\alpha (T,r) = f_{N,M,l_T} (r)$ and so we get the condition
$l(0)=l_0$, $l(T)=l_T$.
Moreover, we get $\zeta(0)=\zeta (T)=0$.

As the next step, we have to solve the main equation \eqref{main_eq_chi}.
We assume that $\vec\chi$ does not depend on $\theta$, i.e.
$\vec\chi = \chi_r (t,r)\, \hat r + \chi_\theta (t,r)\, \hat\theta$.
Equation \eqref{main_eq_chi} then becomes
\beqa
\frac{2mr}{q} \frac{\partial\alpha}{\partial t}
- \chi_r \left(a+2r \frac{\partial\alpha}{\partial r}\right)
- r \frac{\partial \chi_r}{\partial r} \alpha = 0,
\label{r_eq_chi}
\eeqa
and Eq. \eqref{main_eq_phi} becomes
\beqa
\Phi &=& - \frac{q}{2m} \left(\chi_r^2 + \chi_\theta^2\right) + \frac{\hbar^2}{2
  m q \alpha} \left(\frac{1}{r} \frac{\partial\alpha}{\partial r} +
\frac{\partial^2\alpha}{\partial r^2}\right).
\eeqa
A solution of Eq. \eqref{main_eq_chi} is given by
\beq
\chi_r (t,r) = -\frac{2m}{q} \frac{1}{r \, \alpha^2} \int_r^\infty ds\, s\, \alpha
\frac{\partial \alpha}{\partial t}(t,s).
\label{sol}
\eeq
The solution when $\alpha$ and $\vec\chi$
depend on $\theta$ can be found in Appendix \ref{app_solution}.

For $\alpha$ given by Eq. \eqref{alpha}, we get from Eq. \eqref{sol} that
\begin{eqnarray}
\chi_r (t,r) = -\frac{m}{q} \frac{r l' (t)}{l(t)},
\label{chi_r}
\end{eqnarray}
where the prime indicates a derivative with respect to time. Note that this solution is independent 
of the quantum numbers $N$ and $M$.

The components of the physical fields can be written as
\beqa
B_z (t,r) &=& \frac{1}{r} \frac{\partial (r \chi_\theta)}{\partial r},
\\
E_r (t,r) &=&
-\frac{\partial}{\partial r}\Phi
- \frac{\partial \chi_r}{\partial t}, \;
E_\theta (t,r)= -\frac{\partial \chi_\theta}{\partial t}.
\eeqa
We want a uniform magnetic field and
a constant radial electric field $\left(E_r = \frac{m \omega_{z}^2}{2 q} r \right)$ during the whole process.
To achieve a uniform magnetic field $B_z = B_z (t)$ we set
\beq
\chi_\theta (t,r) = \frac{r}{2} B_z(t) + \frac{g(t)}{r},
\eeq
with an arbitrary function $g$.
The electric field component $E_r$ is now
\beqa
E_r &=& \frac{M^2 \hbar^2 - q^2 g(t)^2}{m q r^3}
\nonumber\\
& &+ 
 \frac{r}{4mq}\left(q^2 B_z^2 - \frac{\hbar^2}{l(t)^4}+\frac{4 m^2 l''(t)}{l(t)} \right).
\label{Er}
\eeqa
The demand $E_r = \frac{m \omega_{z}^2}{2 q} r$ leads to
the choice 
$g(t) = - M \hbar/q$
and
\begin{equation}
B_z(t) = \frac{\sqrt{\hbar^2 - 4m^2l(t)^3l''(t)+2 m^2 \omega_{z}^2 l(t)^4}}{q l(t)^2},
\label{Bz}
\end{equation}
where $q B_z (t) > 0$ is assumed. The electric field components are finally
\begin{eqnarray}
E_r = \frac{m \omega_{z}^2}{2 q} r, \; E_\theta = -\frac{r}{2} B_z'(t).
\label{E_total}
\end{eqnarray}
Now, we have
\beq
\chi (t, r) = \left[\frac{r}{2} B_z(t) - \frac{M \hbar}{q r}\right] \hat\theta
-\frac{m}{q} \frac{r l' (t)}{l(t)} \hat r .
\label{chi}
\eeq
Following from Eq. \eqref{bound_chi}, we get the boundary conditions for $\vec\chi$
\begin{eqnarray}
\vec \chi (0,r) = \vec A_0 - \frac{\hbar}{q} \nabla \beta_0 = 
\left(\frac{r B_{0}}{2} - \frac{\hbar M}{q r} \right) \hat\theta,
\nonumber\\
\vec \chi (T,r) = \vec A_T - \frac{\hbar}{q} \nabla \beta_T = 
\left(\frac{r B_{T}}{2} - \frac{\hbar M}{q r} \right) \hat\theta.
\end{eqnarray}
With the $\chi$ given in Eq. \eqref{chi}, this is fulfilled if
$l'(0)= 0$, $l'(T)= 0$,
$B_z(0)=B_0$, and $B_z(T)=B_T$. 
To fulfill the last two conditions, we have to demand
$l''(0)=0$,
$l''(T)=0$.

The boundary conditions of $\Phi$ are fulfilled if the conditions
\eqref{bound_beta} are satisfied, i.e. if
\beqa
\frac{\partial\zeta}{\partial t} (0) = -\frac{1}{\hbar} \mathcal{E}_0,\;
\frac{\partial\zeta}{\partial t} (T) = -\frac{1}{\hbar} \mathcal{E}_T.
\eeqa
A simple choice of $\zeta$ may be a polynomial of degree $3$ that obeys all of
the boundary conditions on $\zeta$. Note that the magnetic and the electric
fields do not depend on the choice of the time dependent global phase $\zeta(t)$.

An additional boundary condition on \(l(t)\) can be derived by enforcing that
the electric field is continuous at $t=0$ and $t=T$, see the
conditions \eqref{bound_electric}.
This requires
$B_z'(0)=0$ and
$B_z'(T)=0$.
Differentiating the expression \eqref{Bz} for \(B_z(t)\) with respect to time gives
\begin{equation}
B_z'(t) = -\frac{2\Bigl(l'(t)\left[\hbar^2 - m^2 l(t)^3 l''(t) \right] 
+ m^2 l(t)^4 l'''(t) \Bigr)}{ql^3(t) \sqrt{\hbar^2 - 4m^2 l(t)^3l''(t)+2m^2 \omega_{z}^2 l(t)^4}}.
\end{equation}
Noting the boundary conditions on \(l\) already derived,
this requires, in addition, that
$l'''(0)=0$, 
$l'''(T)=0$.

In summary, the boundary conditions for \(l(t)\) are
\begin{align}
l(0)=l_0& =\sqrt{\frac{\hbar}{2 m \widetilde{\omega}\left(0\right)}}, \;
l'(0) = l''(0)=l'''(0) = 0  ,\nonumber\\
l(T)=l_T& =\sqrt{\frac{\hbar}{2 m \widetilde{\omega}\left(T\right)}} , \;
l'(T)=l''(T)=l'''(T)=0.
\label{l_boundary}
\end{align}
These conditions are independent
of the quantum numbers \(N\) and \(M\).

If $l(t)$ satisfies these boundary conditions, 
the corresponding magnetic field and electric field are given by Eqs. \eqref{Bz} and
\eqref{E_total}, and fulfill $\nabla\cdot\vec E = 0$ and $\nabla \times \vec B = 0$.
They are also independent of the quantum numbers \(N\) and \(M\).
If the system starts in the corresponding eigenstate and if these fields are
implemented, then the system will end with fidelity $1$ in the final state.
We will show  that these schemes which do not change the quantum
number could be  alternatively derived using an invariant based approach.
\subsection{Invariant based approach}
A Lewis-Riesenfeld invariant is a Hermitian
operator $I(t)$ fulfilling
\beqa
\frac{\partial}{\partial t} I(t) = \frac{i}{\hbar}\left[ I(t), H(t) \right]_-,
\eeqa
where $H(t)$ is the Hamiltonian for the system.
If we disregard again the $z$ dependent part, 
we find the following invariant for the Hamiltonian \eqref{LRHam},
\beqa
I(t) &=& -\frac{l(t)^2}{\hbar^2} \Delta -2ml'(t)l(t) \left(\frac{\hbar}{i} \frac{\partial}{\partial r}\right) r
\nonumber\\
&& + \left(m^2 l'(t)^2 + \frac{\hbar^2}{4 l(t)^2}\right) r^2,
\eeqa
where the function $l(t)$ has to be a solution of the following Ermakov-like equation
\beq
4m^{2}\frac{l''(t)}{l(t)}+4 m^{2}\widetilde{\omega}(t)^2-\frac{\hbar^{2}}{l(t)^{4}}=0.
\label{ermakovok}
\eeq
(In  \cite{LR69}, the case $\widetilde{\omega}=\omega$ was examined and an
invariant was constructed. Eigenstates of this invariant which are
simultaneously eigenstates of $L_z$ were also
constructed indirectly.) An explicit expression of the
eigenstates of $I$ is
\beqa
\lefteqn{\Gamma_{N,M}(t,r,\theta)=\frac{1}{\sqrt{2 \pi}}\sqrt{\frac{N!}{(N+|M|)!}}\frac{1}{l(t)}\left[\frac{r}{\sqrt{2}l(t)}\right]^{|M|}}
&& 
\nonumber\\
&& \times \fexp{-\frac{r^2}{4l(t)^2}}
L^{|M|}_{N}\left(\frac{r^2}{2l(t)^2}\right) \fexp{i M\theta}
 \nonumber\\
&& \times \fexp{\frac{i m l'(t)}{2 \hbar l} r^2},
\label{invstates}
\eeqa
where $N, M \in \bZ$ and $l(t)$ is a solution of Eq. \eqref{ermakovok}.
The corresponding eigenvalue of $I$ is $(2 N+|M|+1)\hbar^2$ and the corresponding
eigenvalue of $L_z$ is $M\hbar$. A more general invariant (which allows for a time dependent mass) was described in \cite{fiore2011}.
An eigenstate of a Lewis-Riesenfeld invariant is a solution to the Schr\"odinger equation for $H(t)$ up to a time dependent phase $\Pi_{N,M}(t)$ \cite{LR69}, which is here given by
\begin{equation}
\Pi_{N,M}(t)=\int_0^t dt' \left( M \omega(t')-\frac{(N+1)\hbar}{2 m
  l(t')^2}\right).
\label{lrphases}
\end{equation}
The idea is to do inverse-engineering by demanding that the system follows the
state $\Psi(t, r,\theta)=\Gamma_{N,M}(t,r,\theta) e^{i \Pi_{N,M}(t)}$.
First, we choose an auxiliary function $l(t)$ (which has to fulfill different
conditions at initial and final time, see below) and then we get
$\widetilde\omega(t)$ from Eq. \eqref{ermakovok}.
At initial and final time the eigenstates of the invariant should coincide with the eigenstates
of the Hamiltonian, i.e. $[I(0),H(0)]_-=0= [I(T),H(T)]_-$. Therefore, we have to impose the following
boundary conditions for the auxiliary function $l(t)$:
\begin{align}
l(0)&=l_0=\sqrt{\frac{\hbar}{2 m \widetilde{\omega}\left(0\right)}}, \;&
l(T)&=l_T=\sqrt{\frac{\hbar}{2 m \widetilde{\omega}\left(T\right)}},
\nonumber\\
l'(0)&=0, \;& l'(T)&=0.
\end{align}
From these boundary conditions and Eq. \eqref{ermakovok}, it also follows
that $l''(0)=0$ and $l''(T)=0$.

An additional boundary condition on \(l(t)\) can be derived by enforcing that
the electric field is continuous at $t=0$
and $t=T$. This leads to 
$B_z'(0) = B_z'(T)=0$ or
$l'''(0)=l'''(T)=0$.
The complete list of boundary conditions is equivalent to Eq. \eqref{l_boundary} above.
As already mentioned, we can now design first an auxiliary function $l(t)$ fulfilling the above
boundary conditions and then calculate $\widetilde\omega(t)$ and hence
the magnetic field strength $B_z(t)$ from Eq. \eqref{ermakovok}. The resulting
formula for $B_z(t)$ is the same as Eq. \eqref{Bz} above.

Summarizing, the streamlined fast-forward formalism and the invariant based
approach provide two ways to find the same boundary conditions
for the auxiliary function $l(t)$ in this setting. So, for varying the magnetic
field strength, both formalisms are equivalent as they
both require that one chooses the auxiliary function $l(t)$ fulfilling
these boundary conditions and then the corresponding physical potentials can
be calculated in the same way. In the following we look at a numerical example of this
procedure.
%
%
\begin{figure}[t]
\begin{center}
\includegraphics[width=0.8\linewidth]{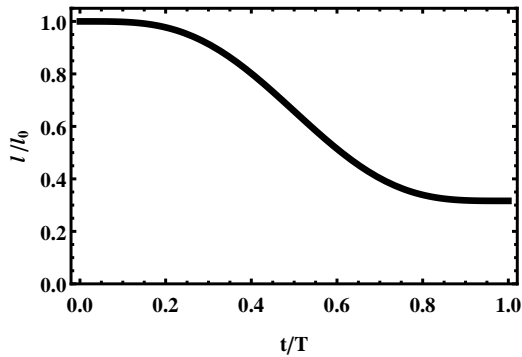}
\caption{\label{fig_1}Auxiliary function $l(t)$ versus $t$.}
\end{center}
\end{figure}
\begin{figure}[t]
\begin{center}
\includegraphics[width=0.8\linewidth]{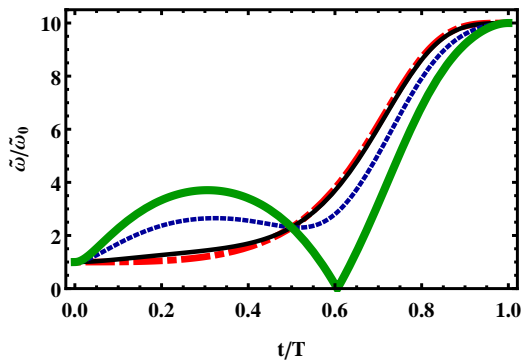}
\caption{\label{fig_2} (Color online) Frequency ratio $\widetilde\omega (t)/\widetilde\omega(0)$ versus $t$;
$\mu \to \infty$ (red, dashed-dotted line), $\mu=3$ (black, solid line),
 $\mu=1$ (blue, dotted line), $\mu=0.672$ (green, thick, solid line).}
\end{center}
\end{figure}
\begin{figure}[t]
\begin{center}
\includegraphics[width=0.8\linewidth]{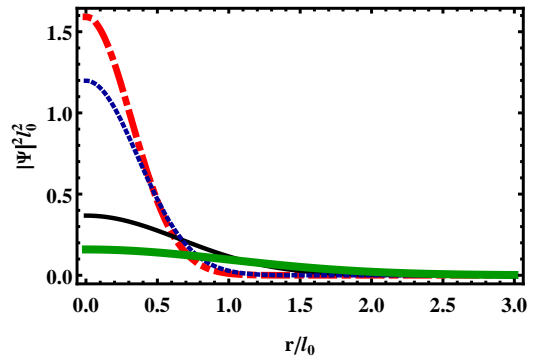}
\caption{\label{fig_3} (Color online) Time evolution of $\fabsq{\Psi(t,r)}$ with $N=M=0$;
$t=0$ (green, thick, solid line),
$t=T/2$ (black, thin, solid line), $t=3T/4$ (blue, dotted line),
$t=T$ (red, dashed-dotted line).}
\end{center}
\end{figure}
%
%
%
\subsection{Numerical Example}
Let us first set $l (t) = l_0 \lambda (\tau)$ where $\tau = t/T$.
From the above formalism, the time dependence of $\widetilde\omega$ follows as
\beq
\widetilde{\omega} (t) = \widetilde{\omega} (0) \frac{1}{\lambda(\tau)^2}
\sqrt{1 - \frac{\lambda(\tau)^3 \lambda''(\tau)}{\mu^2}},
\eeq
where $\mu = T \widetilde{\omega}(0)$.
$\mu$ can be seen as the final time in units of $1/\widetilde{\omega}(0)$.
Therefore, decreasing $\mu$ corresponds to decreasing the total time $T$ of the process,
with fixed $\widetilde\omega (0)$ (i.e. fixed $\omega_0 =\frac{q B_z(0)}{2m}>0$ and fixed $\omega_z$).
The limit $\mu\to\infty$
would correspond to the adiabatic limit where we get
$\widetilde{\omega} (t)/\widetilde{\omega} (0) \to \frac{1}{\lambda^2(\tau)}$.

The corresponding magnetic field would then be given by Eq. \eqref{Bz} or in
dimensionless variables
\begin{eqnarray}
B_z (t) = \frac{\hbar}{q l_0^2} \frac{1}{\lambda(\tau)^2}
\left[1 - \frac{\lambda(\tau)^3 \lambda''(\tau)}{\mu^2} +
  \frac{\nu^2}{2-\nu^2} \lambda(\tau)^4\right]^{1/2}
\end{eqnarray}
where $\nu = \frac{\omega_z}{\omega_0}$ is the ratio between the two initial
frequencies. This parameter $\nu$ is independent of the total time $T$.
We want to have a trap setting at initial and final time,
i.e. $\widetilde{\omega}(0)^2$ and $\widetilde{\omega}(T)^2$ should be positive. 
From this, $\nu$ must be in the range
$0 \le \nu < \sqrt{2} \mbox{min}\{1, \omega_T/\omega_0\}$ where
$\omega_T = \frac{q B_z(T)}{2m}>0$. 

Assuming a polynomial form of $\lambda(\tau)$ and using the above conditions,
$\lambda(\tau)$ can be expressed as
\begin{eqnarray}
\lambda (\tau)&=& 1 -20\left(l_T/l_0 -1\right) \tau^7 +
70 \left(l_T/l_0 - 1\right)\tau^6
\nonumber\\
&& - 84 \left(l_T/l_0 - 1\right)\tau^5 +35 \left(l_T/l_0 - 1\right)\tau^4.
 \label{eql}
\end{eqnarray}
The final value of the magnetic field is chosen in this example such that
$\widetilde\omega (T)/\widetilde\omega (0) = c = 10$ (i.e. $l_T = l_0/\sqrt{10}$).
The ratio between initial and final magnetic field is then
\begin{eqnarray*}
\frac{B_z(T)}{B_z(0)} =
\sqrt{c^2\left(1-\frac{\nu^2}{2}\right)+\frac{\nu^2}{2}}.
\end{eqnarray*} 
Fig. \ref{fig_1} is the corresponding plot of $l(t)$.

Fig. \ref{fig_2} shows $\widetilde{\omega}(t)$ for different values
of $\mu$. For $\mu \approx 0.672$ (green, thick, solid line)
$\widetilde{\omega} (t)^2 > 0 $ is no longer fulfilled for
all times. The requirement that $B_z (t) \in \mathbb{R}$ for all times results
in a type of quantum speed limit of the form
\begin{eqnarray}
\mu \geq \max_{\tau \in [0,1]} \left[\frac{\lambda(\tau)^3 \lambda''(\tau)}{1+\nu^2 (2-\nu^2)^{-1} \lambda(\tau)^4} \right]^{1/2}.
\end{eqnarray}
As an example, the wavefunction at different times with $N=M=0$ can be seen in Fig. \ref{fig_3}.
The shown time evolution is independent of $\mu,\nu$ and depends only on the chosen form of $l(t)$.
\subsection{Superposition}
The electric and magnetic fields derived in the previous subsection are
independent of the quantum numbers $N$ and $M$. Therefore,
the fields can be also applied to a superposition of different eigenstates
with initial magnetic field $B_0$ and they will
produce a superposition of eigenstates with final magnetic field $B_T$ with
the same populations as initially. Let us assume an initial wavefunction of
the form
\begin{eqnarray}
\Psi (0,r,\theta) = \sum_{N,M} c_{N,M} \Gamma_{N,M} (0,r,\theta),
\end{eqnarray}
where $\Gamma_{N,M} (t,r,\theta)$ are the eigenfunctions of the invariant
given in Eq. \eqref{invstates} and $c_{N,M}$ are (constant) complex coefficients.
Then it follows that the state at final time will be
\begin{eqnarray}
\Psi (T,r,\theta) = \sum_{N,M} c_{N,M} e^{i \Pi_{N,M}(T)} \Gamma_{N,M} (T,r,\theta),
\end{eqnarray}
where $\Pi_{N,M}$ is given in Eq. \eqref{lrphases}, 
so the populations in the different eigenstates will be the same as initially, i.e.
$\fabsq{c_{N,M} e^{i \Pi_{N,M (t)}}} = \fabsq{c_{N,M}}$.
%
\begin{figure}
\begin{center}
\includegraphics[width=0.8\linewidth]{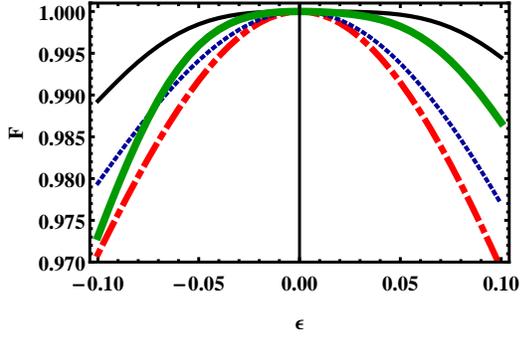}
\caption{\label{fig_4} (Color online) Fidelity $F$ versus error $\epsilon$; 
$\mu=1, \nu=0.1$ (blue, dotted line);
$\mu=1, \nu=1$ (red, dashed-dotted line);
$\mu=3, \nu=0.1$ (black, thin, solid line);
$\mu=3, \nu=1$ (green, thick, solid line).}
\end{center}
\end{figure}
%
\begin{figure}
\begin{center}
\includegraphics[width=0.8\linewidth]{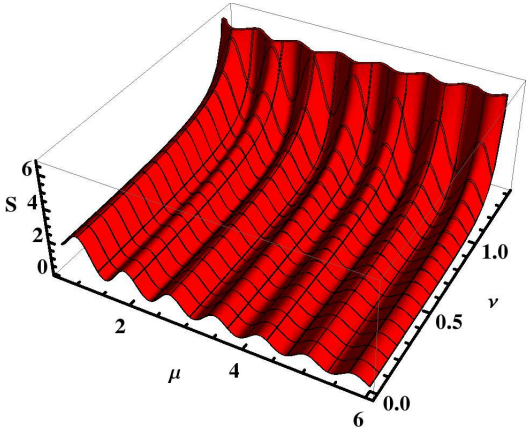}
\caption{\label{fig_5} (Color online) Systematic error sensitivity $S$ versus $\mu$ and $\nu$.}
\end{center}
\end{figure}
%
%
\subsection{Stability}
It is important that the scheme is not only fast but also stable
concerning errors in the implementation.
We want to examine the stability of the protocol if there is a systematic error in the 
magnetic field $B_z(t)$. We assume that the magnetic field is
correctly implemented before and after the process, for $t\le 0$ and $t\ge T$.
Nevertheless, during the change of the magnetic field for  for $0<t<T$,
we assume that an inaccurate magnetic field $\mathcal{B}_\epsilon (t) = B_z (t)(1+\epsilon)$ is
implemented, where $ B_z (t)$ is the correct one
and $\epsilon$ a small relative systematic error which is
unknown but constant. 

We will examine the final fidelity as a function of $\epsilon$ for 
$N=M=0$. The initial state is still $\Psi_\epsilon (0) = \psi_{N=0,M=0,l_0}$.
The solution of the Schr\"odinger equation is then still given by $\Psi_\epsilon
(t) = \Gamma_{0,0} (t) e^{i \Pi_{0,0}(t)}$
(see Eq. \eqref{invstates}) with $l(t)$ replaced by $\ell_\epsilon (t)$, a solution of
\beq
4m^{2}\frac{\ell_\epsilon ''(t)}{\ell_\epsilon (t)}+4 m^{2}
\left[ \left(\frac{q \mathcal{B}_\epsilon (t)}{2m}\right)^2 - \frac{\omega_z^2}{2}\right]-\frac{\hbar^{2}}{\ell_\epsilon (t)^{4}}=0,
\label{ermakov}
\eeq
with $\ell_\epsilon (0)=l_0$ and $\ell_\epsilon '(0)=0$. The fidelity at $t=T$ is now
\begin{eqnarray}
F &=& \fabs{\la \Psi(T) | \Psi_{\epsilon}(T) \ra}
\nonumber\\
&=& \frac{2 l (T) \ell_\epsilon (T)}{\sqrt{(l(T)^2 + \ell_\epsilon (T)^2)^2 + \frac{4m}{\hbar^2} l(T)^4 \ell_\epsilon (T)^2 \ell_\epsilon '(T)^2}}.
\label{fidelity}
\end{eqnarray}
Let $l(t)$ be given again as in Eq. (\ref{eql}). We once again fix
$\widetilde\omega (T) / \widetilde\omega (0) = c = 10$, noting that the magnetic field
is assumed to be error-free at the initial and final time.
With these values fixed, the fidelity $F$ only depends on $\mu$, $\nu$ and
$\epsilon$. Note $\nu$ must be in the range $0 \le \nu < \sqrt{2}$.
The fidelity $F$ for different combinations of $\mu$ and $\nu$ versus $\epsilon$
is shown in Fig. \ref{fig_4}. One still gets a high fidelity even if there is a small, systematic error in the
implementation of the magnetic field during the scheme. The scheme
is, in some range,  stable concerning this type of systematic error.

A sensitivity $S$ of the scheme versus this systematic error can be defined as
the negative curvature of the fidelity at $\epsilon=0$, i.e.
$S = -\frac{\partial^2 F}{\partial \epsilon^2}\big|_{\epsilon=0}$. This
sensitivity $S$ versus $\mu$ and $\nu$ is shown in Fig. \ref{fig_5}. 
The sensitivity is increasing with increasing ratio $\nu$ for fixed $\mu$. For
fixed $\nu$ the sensitivity shows an 
oscillating behavior with increasing $\mu$. The sensitivity for arbitrary $N$
and $M$ is treated in Appendix \ref{app_fidelity}.

In order to have even more stability against systematic error in the magnetic
field one could design a different $l(t)$ which minimizes the
sensitivity $S$ and still fulfills the necessary boundary conditions (a similar strategy could also be applied
to other types of systematic errors or random errors). As shown
in Appendix \ref{app_fidelity}, it would be sufficient to minimize $S$ only
for $N,M = 0$.
%
%
%
%
\section{Discussion \label{Discussion}}
We have put forward shortcuts to adiabaticity for a charged particle in an electromagnetic
field  focusing on a change of the radial spread in a Penning trap by modifying the magnetic field
intensity. 
Two methods have been used for this: a 
streamlined version of the fast-forward formalism for an electromagnetic field, and 
an invariant based procedure.
We have shown their equivalence for this operation. 
In general the fast-forward formalism presented in this paper could be applied to 
other tasks for which the invariant approach is not well suited, such as transformations 
for individual states \cite{torrontegui2012}.   
We also found that a type of quantum speed limit applies.
Moreover, we have examined the scheme in the case of a systematic
error in the magnetic field and shown its stability. 
%
%
%

\begin{acknowledgments}
We are grateful to David Rea for useful discussion and commenting on the manuscript.
This work was supported by the Basque Country Government (Grant No. IT472-10), Ministerio de Econom\'\i a y Competitividad (Grant No. FIS2012-36673-C03-01), and the program UFI 11/55 of the Basque Country University.

\end{acknowledgments}

\begin{appendix}
\section{Boundary conditions for $\Phi(t,\vec r)$ \label{app_boundary}}
Because $\psi_0 (\vec r) \equiv \alpha_0(\vec r) e^{i \beta_0(\vec r)}$ should
be an energy eigenvector of the Hamiltonian $H_0$ with eigenvalue
$\mathcal{E}_0$, it follows from the real part of the corresponding stationary Schr\"odinger
equation that
\beqa
0 = -\frac{\hbar^2}{2m} \Delta \alpha_0 + \frac{1}{2m} \left(q\vec A_0 - \hbar
\nabla \beta_0\right)^2 \alpha_0 + \left(q \phi - \mathcal{E}_0\right) \alpha_0, 
\nonumber\\
\label{aeq_1}
\eeqa
and from the imaginary part that
\beqa
0
= \frac{\hbar}{2m}\nabla\left(q\vec A_0 - \hbar\nabla \beta_0\right) \alpha_0
+ \frac{\hbar}{m}\left(q\vec A_0 - \hbar\nabla \beta_0\right) \nabla \alpha_0.
\nonumber\\
\label{aeq_2}
\eeqa
Eq. \eqref{main_eq_chi} at initial time becomes
\beqa
\frac{1}{q}\frac{\partial\alpha}{\partial t} (0, \vec r)
&=& \frac{1}{2m}(\nabla\vec\chi_0) \alpha_0
+ \frac{1}{m} \vec\chi_0 \nabla \alpha_0 = 0,
\eeqa
because of Eq. \eqref{aeq_2}, and $\vec \chi_0 = \vec A_0 - \frac{\hbar}{q}\nabla \beta_0$.

Eq. \eqref{main_eq_phi} at initial time becomes
\beqa
\Phi (0, \vec r) &=& \frac{\hbar^2}{2 m q \alpha_0} \Delta \alpha_0 - \frac{q}{2m} \vec\chi_0^2 
= \phi_0 - \frac{\mathcal{E}_0}{q}
\eeqa
because of Eq. \eqref{aeq_1}.
Similar calculations also apply to the final time.
\section{Solution of the main equations in polar coordinates \label{app_solution}}
We assume that $\alpha (t,r,\theta)$ is given.
We set
\beqa
\vec\chi = \chi_r (t,r,\theta)\, \hat r + \chi_\theta (t,r,\theta)\,
\hat\theta.
\eeqa
The main equation  \eqref{main_eq_chi} now becomes
\beqa
\frac{2mr}{q} \frac{\partial\alpha}{\partial t} - 2 \chi_\theta
  \frac{\partial \alpha}{\partial\theta}
- \frac{\partial \chi_\theta}{\partial\theta} \alpha
- \chi_r \left(\alpha+2r \frac{\partial\alpha}{\partial r}\right)
\nonumber\\
- r \frac{\partial \chi_r}{\partial r} \alpha = 0
\label{ap_eq_chi}
\eeqa
and Eq. \eqref{main_eq_phi} becomes
\beqa
\Phi &=& - \frac{q}{2m} \left(\chi_r^2 + \chi_\theta^2\right)
\nonumber\\
&&+ \frac{\hbar^2}{2
  m q \alpha} \left(\frac{1}{r^2} \frac{\partial^2\alpha}{\partial\theta^2} +
\frac{1}{r} \frac{\partial\alpha}{\partial r} +
\frac{\partial^2\alpha}{\partial r^2}\right).
\label{ap_eq_phi}
\eeqa
A solution of the main equation \eqref{ap_eq_chi} for $\chi_r$ in
terms of $w=\alpha^2$ and $\chi_\theta$ can be written down,
\beqa
\lefteqn{\chi_r = - \frac{1}{r w} 
\int_r^\infty ds\,
\Bigg[
\frac{m s}{q} \frac{\partial w}{\partial t} (t,s,\theta)}&& 
\nonumber\\
&& - \chi_\theta(t,s,\theta)
\frac{\partial w}{\partial\theta}(t,s,\theta) - w(t,s,\theta)
\frac{\partial \chi_\theta}{\partial\theta}(t,s,\theta)\Bigg].
\nonumber\\
\eeqa
Thus a function $\chi_\theta$ can be chosen to determine (together with the
chosen $\alpha$) the function $\chi_r$. 

Alternatively, a solution of the main equation for $\chi_\theta$ in terms of
$a$ and $\chi_r$ is given by
\beqa
\lefteqn{\chi_\theta=-\frac{f_\theta (r,t)}{a} + \frac{1}{a} \int_0^\theta d\rho\,
\Bigg\{\ 
\frac{m r}{q} \frac{\partial a^2}{\partial t} (r,\rho,t)} && 
\nonumber\\
&&- r \chi_r(r,\rho,t) \frac{\partial a^2}{\partial r}(r,\rho,t)
\nonumber\\
&&- a^2(r,\rho,t) \left[\chi_r(r,\rho,t) + r \frac{\partial
  \chi_r}{\partial r}(r,\rho,t)\right]\Bigg\},
\eeqa
where $\chi_\theta (r, \theta + 2\pi, t) =
\chi_\theta (r, \theta, t)$.
\section{Fidelity and sensitivity for arbitrary $N$ and $M$ \label{app_fidelity}}
For arbitrary $N$ and $M$ we get for the fidelity $F_{N,M}$, by using \cite{gradstein},
\begin{eqnarray}
F_{N,M} &=& \fabs{\la \Psi(T) | \Psi_\epsilon(T) \ra}\nonumber\\
&=& Q^{1+\fabs{M}} \fabs{P_N^{(\fabs{M},0)} (1-2 Q^2)},
\end{eqnarray}
where $P$ are Jacobi's polynomials and
\begin{eqnarray}
Q = \frac{2 l (T) \ell_\epsilon (T)}{\sqrt{(l(T)^2 + \ell_\epsilon (T)^2)^2 + \frac{4m}{\hbar^2} l(T)^4 \ell_\epsilon(T)^2 \ell_\epsilon'(T)^2}}.
\end{eqnarray}
The result is valid for an arbitrary function $l(t)$.
Note that $F_{N=0,M=0}= Q = F$, this is the special case given in Eq. \eqref{fidelity}.

The general sensitivity $S_{N,M} = -\left.\frac{\partial^2 F_{N,M}}{\partial \epsilon^2}\right|_{\epsilon=0}$ is
\beq
S_{N,M} = \left.\frac{\partial F_{N,M}}{\partial Q}\right|_{Q=1} \times S_{N=0,M=0},
\eeq
where $S_{N=0,M=0}=S$ is the special case discussed in the main text.
The factor $\left.\frac{\partial F_{N,M}}{\partial Q}\right|_{Q=1}$ only depends on $N$ and $M$
and is independent of the chosen $l(t)$. 
\end{appendix}

\end{document}